\documentclass[conference]{IEEEtran}
\IEEEoverridecommandlockouts
\usepackage{cite}
\usepackage{amsmath,amssymb,amsfonts}
\usepackage{algorithmic}
\usepackage{graphicx}
\usepackage{textcomp}
\usepackage{xcolor}
\def\BibTeX{{\rm B\kern-.05em{\sc i\kern-.025em b}\kern-.08em
    T\kern-.1667em\lower.7ex\hbox{E}\kern-.125emX}}
\begin{document}

\title{Emo-CNN for Perceiving Stress from Audio Signals: A Brain Chemistry Approach\\
\thanks{}
}

\author{\IEEEauthorblockN{1\textsuperscript{st} Anup Anand Deshmukh}
\IEEEauthorblockA{\textit{University of Waterloo, Canada} \\
\textit{\emph{aa2deshmukh@uwaterloo.ca}}\\
}
\and
\IEEEauthorblockN{2\textsuperscript{nd} Catherine Soladie}
\IEEEauthorblockA{\textit{Centralesupelec, France} \\
\textit{\emph{Catherine.Soladie@supelec.fr}}\\
}
\and
\IEEEauthorblockN{3\textsuperscript{rd} Renaud Seguier}
\IEEEauthorblockA{\textit{Centralesupelec, France} \\
\textit{\emph{renaud.seguier@centralesupelec.fr}}\\
}
}

\maketitle

\begin{abstract}

Emotion plays a key role in many applications like healthcare, to gather patients’ emotional behavior. While typical ASR (Automated Speech Recognition) problems focus on “what was said”, it is equally important to understand “how it was said.” There are certain emotions which are given more importance due to their effectiveness in understanding human feelings. In this paper, we propose an approach that models human stress from audio signals. The research challenge in speech emotion detection is finding the appropriate set of acoustic features corresponding to an emotion. Another difficulty lies in defining the very meaning of stress and being able to categorize it in a precise manner. Supervised Machine Learning models, including state of the art Deep Learning classification methods, rely on the availability of clean and labelled data. One of the problems in affective computation and emotion detection is the limited amount of annotated data of stress. The existing labelled stress emotion datasets are highly subjective to the perception of the annotator. \\

We address the first issue of feature selection by exploiting the use of traditional MFCC features in Convolutional Neural Network. Our proposed Emo-CNN architecture treats speech representations in a manner similar to how CNN’s treat images in a computer vision problem. Our experiments show that Emo-CNN consistently and significantly outperforms the popular existing methods over multiple datasets. It achieves 90.2\% categorical accuracy on the Emo-DB dataset. We claim that Emo-CNN is robust to speaker variations and environmental distortions. The proposed approach achieves 85.5\% speaker-dependant categorical accuracy for SAVEE dataset, beating the existing CNN based approach by 10.2\%.  \\

To tackle the second and the more significant problem of subjectivity in stress labels, we use Lovheim's cube, which is a 3-dimensional projection of emotions. Monoamine Neurotransmitters are a type of chemical messengers in the brain that transmit signals on perceiving emotions. The cube aims at explaining the relationship between these neurotransmitters and the positions of emotions in 3D space. The learnt emotion representations from the Emo-CNN are mapped to the cube using three component PCA (Principal Component Analysis) which is then used to model human stress. This proposed approach not only circumvents the need for labelled stress data but also complies with the psychological theory of emotions given by Lovheim's cube. We believe that this work is the first step towards creating a connection between Artificial Intelligence and the chemistry of human emotions.
\end{abstract}


\section{Emo-CNN}

It has become increasingly important to understand human emotions especially stress in many healthcare applications. The ultimate goal of this work is to build a model capable of classifying stress and non-stress audio signals. Within the first step, we use CNN which is trained for a  classification task over seven emotions. Those seven emotions are: Angry, Boredom, Disgust, Fear, Happy, Neutral and Sad. The feature set used for each audio signal is $199 \times 39$ dimensional.

In order to investigate the performance of Emo-CNN, we compare our algorithm with SBS+SVM \cite{semwal2017automatic}, MSF+LDA \cite{wu2011automatic} and Semi-CNN \cite{huang2014speech} methods on Emo-DB dataset. The Emo-DB dataset has 535 clips from 10 actors, 429 of which are used for training. Table \ref{tab:res1} shows the improvement of Emo-CNN over aforementioned methods with regards to the classification accuracy.

\begin{table}[htbp] \label{tab:res1}
\begin{center}
    \caption{Entries describe the comparison between the Emo-CNN and other approaches }
  \begin{tabular}{|c|c|c|c|}
  \hline
    Model & Categorical Accuracy\\
    \hline
    SBS+SVM \cite{semwal2017automatic} & 80\% \\
    MSF+LDA \cite{wu2011automatic} & 85.6\% \\
    Semi-CNN \cite{huang2014speech} & 88.3\% \\
    Emo-CNN & 90.2\% \\
    \hline
\end{tabular}
\label{values}
\end{center}
\end{table}

\section{Lovheim's cube}

\begin{figure}  \label{fig:cube}
\begin{center}
 \includegraphics[scale=1.7]{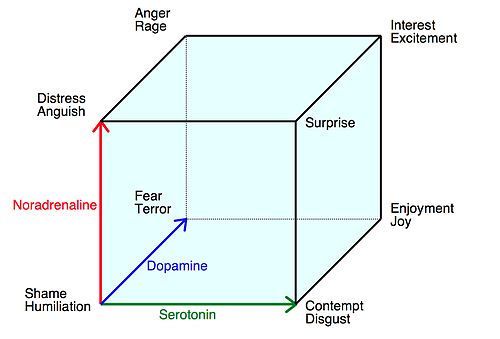}
\caption{Lovheim's cube of emotions \cite{lovheim2012new}}
\end{center}
 \end{figure}

\begin{figure*} \label{fig:main}
\begin{center}
 \includegraphics[scale=0.7]{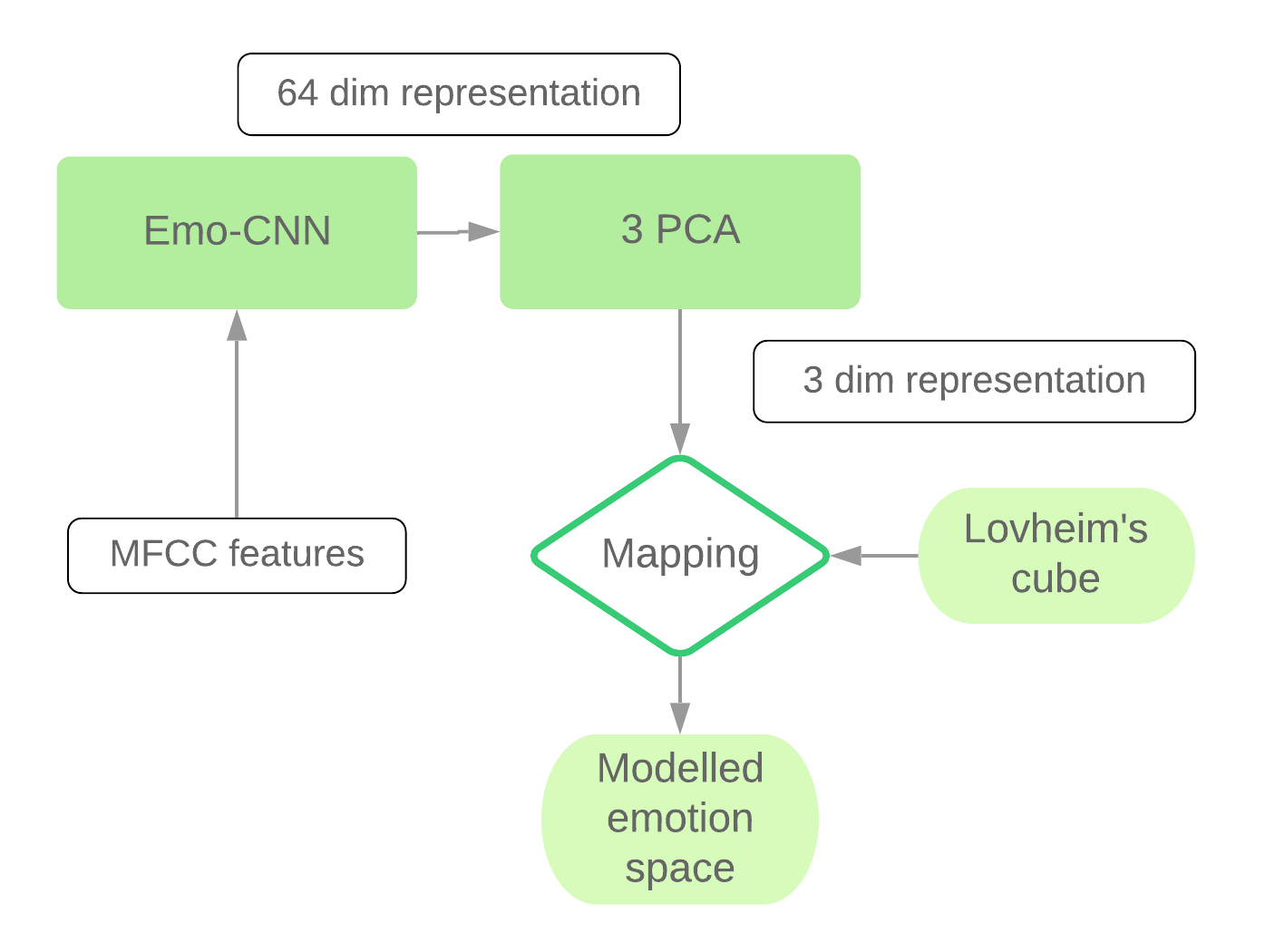}
\caption{The proposed approach which circumvents the subjectivity in stress labels}
\end{center}
\end{figure*}

There have been lot of efforts in defining emotions in a multi-dimensional space. Such models of emotions aim at modelling human emotions by defining where they lie in two or three dimensions. The key idea behind having multiple dimensions is to incorporate the neurophysiological system which causes different affective states in humans.   

To fulfill our final goal of identifying stress from audio signals, we take the help of Lovheim's cube \cite{lovheim2012new}. This cube gives the direct relation between specific combinations of the levels of the signal substances which are produced in our bodies and eight basic emotions. These signal substances are called as Neurotransmitters which are nothing but the messengers transmitting signals across a chemical synapse in our bodies. The figure 1 shows a Lovheim's cube of emotion where three Neurotransmitters, dopamine, noradrenaline and serotonin form the axes of a coordinate system. The eight basic emotions including seven emotions on which our CNN is trained and the emotion - stress (distress), are placed in the eight corners. 

We first take the 64 dimensional representation from the second last layer of Emo-CNN and feed it to 3 PCA. This 3 dimensional representation of audio signal is then mapped onto the Lovheim's cube. The table \ref{tab:res2} shows the mapped values of test audio signals by CNN + 3 PCA. The emotion Happy (Joy) according to the Lovheim's model is produced by the combination of low noradrenaline, high dopamine and high serotonin. Our CNN + 3 PCA model's learnt representation gives the levels of these 3 Neurotransmitters as -2.00, 0.16 and 0.69 resp. From table \ref{tab:res2} we can see that this computational method complies with the theory of Lovheim's cube from psychology. Since our proposed method can model the Lovheim's cube we can now use the 3 dimensional features of audio signals and check their proximity to the stress (Distress) point of the cube. Since Lovheim's cube gives us the relative position of stress from other emotions in 3D space, the proposed approach can easily identify the stressed audio speech without using the labelled stress data. Refer to figure 2.

\begin{table}[htbp]\label{tab:res2}
\caption{Mapping of CNN + 3 PCA on Lovheim's cube}
\begin{center}
\begin{tabular}{|c| c | c | c | c |}
\hline
\textbf{Emotion label} & \multicolumn{3}{|c|}{CNN $+$ 3PCA} \\
\hline
  & Dopamine & Noradrenaline & Serotonin  \\
\hline
Angry & 1.49 & 0.58 & -0.21  \\
\hline
Happy & 0.16 & -2.00 & 0.69   \\
\hline
Fear & 0.19 & -0.68 & -1.77  \\
\hline
Disgust & -0.34 & -0.14 & 0.16  \\
\hline

\end{tabular}
\end{center}
\end{table}

\section{Future Work}
This work shows the potential of Deep Learning models in understanding the chemistry of human emotions. It is very interesting to note that although Emo-CNN was just trained on audio signals and emotion labels, it was also able to model the brain chemistry of these emotions. There is significant amount of research still to be conducted to determine the validity and reliability of this model, particularly in having the generalizable and meaningful mapping of features onto Lovheim's cube. Specifically, next steps would be to have a more precise method to find the proximity of test audio signals to the stress (Distress) point of the cube. 

\bibliography{bib} 
\bibliographystyle{ieeetr}

\end{document}